\documentstyle[aps,multicol,epsfig,amssymb]{revtex}
\begin{document}
\title{Quantum enhanced positioning and clock
synchronization}
\author{V. Giovannetti, S. Lloyd$^*$, and L. Maccone.}
\address{Massachusetts Institute of Technology, Research
Laboratory of Electronics\\
$^*$Corresponding Author: Department of Mechanical Engineering
MIT 3-160,\\ Cambridge, MA 02139, USA.}
\maketitle

\begin{multicols}{2}
{\bf A wide variety of positioning and ranging procedures are based on
repeatedly sending electromagnetic pulses through space and measuring
their time of arrival.  This paper shows that quantum entanglement and
squeezing can be employed to overcome the classical power/bandwidth
limits on these procedures, enhancing their accuracy.  Frequency
entangled pulses could be used to construct quantum positioning
systems (QPS), to perform clock synchronization, or to do ranging
(quantum radar): all of these techniques exhibit a similar enhancement
compared with analogous protocols that use classical light.  Quantum
entanglement and squeezing have been exploited in the context of
interferometry
{\cite{sqinterfer,sqinterfer1,sqinterfer1b,sqinterfer2,sqinterfer3}},
frequency measurements {\cite{wine}}, lithography {\cite{lith}}, and
algorithms {\cite{grover}}. Here, the problem of positioning a party
(say Alice) with respect to a fixed array of reference points will be
analyzed.}

Alice's position may be obtained simply by sending pulses that
originate from her position and measuring the time it takes for each
pulse to reach the reference points.  The time of flight, the speed of
the pulses and the arrangement of the reference points determine her
position.  The accuracy of such a procedure depends on the number of
pulses, their bandwidth and the number of photons per pulse. This
paper shows that by measuring the correlations between the times of
arrival of $M$ pulses which are frequency-entangled, one can in
principle increase the accuracy of such a positioning procedure by a
factor $\sqrt{M}$ as compared to positioning using unentangled pulses
with the same bandwidth.  Moreover, if number-squeezed pulses can be
produced {\cite{yama}}, it is possible to obtain a further increase in
accuracy of $\sqrt{N}$ by employing squeezed pulses of $N$ quanta,
{\it vs.}  employing ``classical'' coherent states with $N$ mean
number of quanta.  Combining entanglement with squeezing gives an
overall enhancement of $\sqrt{MN}$.  In addition, the procedure
exhibits improved security: because the timing information resides in
the entanglement between pulses, it is possible to implement
{\cite{entloss}} quantum cryptographic schemes that do not allow an
eavesdropper to obtain information on the position of Alice.  The
primary drawbacks of this scheme are the difficulty of creating the
requisite entanglement and the sensitivity to loss. On the other hand,
the frequency entanglement allows similar schemes to be highly robust
against pulse broadening due to transit through dispersive media
{\cite{chiao}}.\par

The clock synchronization problem can be treated using the same
method.  In Refs.  {\cite{jozsa}} and {\cite{chuang}} two novel
techniques for clock synchronization using entangled states are
presented. However, the authors of Ref.  {\cite{jozsa}} themselves
point out that the resources needed by their scheme could be used to
perform conventional clock synchronization without entanglement.
Similarly, all the enhancement of {\cite{chuang}} arises from
employing high-frequency atoms which themselves could be used for
clock synchronization to the same degree of accuracy without any
entanglement.  In neither case do these proposals give an obvious
enhancement over classical procedures that use the same
resources. Here, by contrast, it is shown that quantum features such
as entanglement and squeezing can in principle be used to supply a
significant enhancement of the accuracy of clock synchronization as
compared to classical protocols using light of the same frequency and
power. In fact, the clock synchronization can be accomplished by
sending pulses back and forth between the parties whose clocks are to
be synchronized and measuring the times of arrival of the pulses
(Einstein's protocol). In this way synchronization may be treated on
the same basis as positioning and the same accuracy enhancements may
be achieved through entanglement and squeezing. In this paper only the
positioning accuracy enhancement will be addressed in detail.

\begin{figure}[hbt]
\begin{center}\epsfxsize=.6
\hsize\leavevmode\epsffile{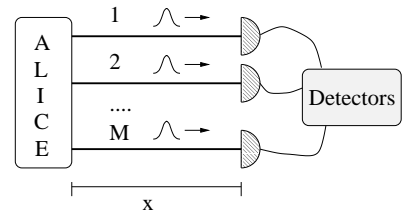} 
\end{center}
\caption{Sketch of the idealized experimental configuration. Alice
sends $M$ light pulses to the $M$ detectors. She averages the times of
arrival $t_i$ of the pulses to recover her unknown position $x$.}
\label{f:localizz}\end{figure}

In order to introduce the formalism, the simple case of position
measurement with classical coherent pulses is now presented.  Since
each dimension can be treated independently, the analysis will be
limited to the one-dimensional case. For the sake of simplicity,
consider the situation in which Alice wants to measure her position
$x$ by sending a pulse to each of $M$ detectors placed in a known
position (refer to Fig. {\ref{f:localizz}}). This can be easily
generalized to different setups, such as the case in which the
detectors are not all in the same location, the case in which only one
detector is employed with $M$ time-separated pulses, the case in which
the pulses originate from the reference points and are measured by
Alice (as in GPS), {\it etc.}  Alice's estimate of her position is
given by $x=c \frac 1M\sum_{i=1}^M t_i$, where $t_i$ is the travel
time of the $i-$th pulse and $c$ is the light speed. The variable
$t_i$ has an intrinsic indeterminacy dependent on the spectral
characteristics and mean number of photons $N$ of the $i$-th
pulse. For example, given a Gaussian pulse of frequency spread
$\Delta\omega$, according to the central limit theorem, $t_i$ cannot
be measured with an accuracy better than $1/(\Delta\omega\sqrt{N})$
since it is estimated at most from $N$ data points ({\it i.e.} the
times of arrival of the single photons, each having an indeterminacy
$1/\Delta\omega$). Thus, if Alice uses $M$ Gaussian pulses of equal
frequency spread, the accuracy in the measurement of the average time
of arrival is
\begin{equation}
\Delta t=\frac{1}{{\Delta\omega}\sqrt{MN}} \;.
\label{deltax}
\end{equation}

Quantum Mechanics allows us to do much better. In order to demonstrate
the gain in accuracy afforded by Quantum Mechanics, it is convenient
to provide first a fully quantum analysis of the problem of
determining the average time of arrival of a set of $M$ classical
pulses, each having mean number of photons $N$. Such a quantum
treatment for a classical problem may seem like overkill, but once the
quantum formalism is presented, the speed-up attainable in the fully
quantum case can be derived directly. In addition, it is important to
verify that no improvement over Eq. (\ref{deltax}) is obtainable using
classical pulses.  The $M$ coherent pulses are described by a state of
the radiation field of the form
\begin{eqnarray} |\Psi\rangle_{cl}\equiv\bigotimes_{i=1}^M
\bigotimes_\omega
\left|\alpha(\phi_\omega\sqrt{N})\right\rangle_i
\;\label{statoclassico},
\end{eqnarray}
where $\phi_\omega$ is the pulses' spectral function,
$|\alpha(\lambda_\omega)\rangle_i$ is a coherent state of amplitude
$\lambda_\omega$ in the mode at frequency $\omega$ directed towards
the $i$-th detector, and $N$ is the mean number of photons in each
pulse. The pulse spectrum $|\phi_\omega|^2$ has been normalized so
that $\int d\omega|\phi_\omega|^2=1$. For detectors with perfect time
resolution, the joint probability for the $i$-th detector to detect
$N_{i}$ photons in the $i$-th pulse at times $t_{i,k}$ is given by
{\cite{tbst}}
\begin{equation} 
p(\{t_{i,k}\}) \propto
\left\langle : \prod_{i=1}^{M} \prod_{k=1}^{N_{i}} 
E_i^{(-)}(t_{i,k})E_i^{(+)}(t_{i,k}) : \right\rangle \;,
\label{timeofarrivalprob}
\end{equation}
where $t_{i,k}$ is the time of arrival of the $k$-th photon in the
$i$-th pulse, shifted by the position of the detectors $t_{i,k}\to
t_{i,k}+x/c$.  The signal field at the position of the $i$-th detector
at time $t$ is given by $E_i^{(-)}(t)\equiv\int d\omega\
a_i^\dag(\omega)\; e^{i\omega t}$ and
$E_i^{(+)}\equiv\left(E_i^{(-)}\right)^\dag$, where $a_i(\omega)$ is
the field annihilator of a quantum of frequency $\omega$ at the $i$-th
detector, which satisfies
$[a_i(\omega),a^\dag_j({\omega'})]=\delta_{ij}\delta(\omega-\omega')$.
The estimation of the ensemble average in
Eq. (\ref{timeofarrivalprob}) on the state $|\Psi\rangle_{cl}$, using
the property $a(\omega')\otimes_\omega|\alpha(\lambda_\omega)\rangle
=\lambda_{\omega'}\otimes_\omega|\alpha(\lambda_\omega)\rangle$, gives
\begin{eqnarray} p(\{t_{i,k}\}) \propto \prod_{i=1}^M 
\prod_{k=1}^{N_{i}} |g(t_{i,k})|^2
\;\label{classicp},
\end{eqnarray}
where $g(t)$ is the Fourier transform 
of the spectral function $\phi_\omega$.  Averaging over the times of
arrival $t_{i,k}$ and over the number of photons $N_{i}$ detected in
each pulse, one has 
\begin{eqnarray} &&\langle t\rangle=\langle
\frac{1}{M}\sum_{i=1}^M\frac 1{N_i}\sum_{k=1}^{N_{i}}t_{i,k}\rangle
= \overline{\tau}
\;\label{classictime}\>;\
\Delta t \gtrsim \frac {\Delta\tau}{\sqrt{MN}}\;,
\end{eqnarray}
with approximate equality for $N\gg1$.  Here $\overline\tau\equiv\int
dt\ t\,|g(t)|^2$ and $\Delta\tau^2\equiv\int dt\
t^2\;|g(t)|^2-{\overline\tau}^2$ are independent of $i$ and $k$ since
all the photons have the same spectrum.  Eq. (\ref{classictime}) is
the generalization of (\ref{deltax}) for non-Gaussian pulses.\par

Quantum light can exhibit phenomena that are not possible classically
such as entanglement and squeezing, which, as will now be seen, can
give significant enhancement for determining the average time of
arrival.  First consider entanglement. The framework just established
allows the direct comparison between frequency entangled pulses and
unentangled ones. For the sake of clarity, consider initially single
photon entangled pulses.\par

Define the ``frequency state'' $|\omega\rangle$ for the
electromagnetic field the state in which all modes are in the vacuum
state, except for the mode at frequency $\omega$ which is populated by
one photon. Thus the state $\int d\omega\;\phi_\omega|\omega\rangle$
represents a single photon wave packet with spectrum
$|\phi_\omega|^2$. Consider the $M$-photon frequency entangled state
given by
\begin{eqnarray} |\Psi\rangle_{en}\equiv\int d\omega
\;\phi_\omega\;|\omega\rangle_1|\omega\rangle_2\cdots|\omega\rangle_M
\;\label{freqmes} ,
\end{eqnarray}
where the ket subscripts indicate the detector each photon is 
traveling to.
Inserting $|\Psi\rangle_{en}$ in Eq. (\ref{timeofarrivalprob}), and
specializing to the case $N_{i}=1$, it follows that 
\begin{eqnarray}
p(t_{1}, \cdots, t_{M}) \propto|g(\sum_{i=1}^Mt_i)|^2
\;\label{probmes}.
\end{eqnarray}
That is, the entanglement in frequency translates into the bunching of
the times of arrival of the photons of different pulses: although
their individual times of arrival are random, the average
$t\equiv\frac 1M\sum_{i=1}^Mt_i$ of these times is highly peaked. (The
measurement of $t$ follows from the correlations in the times of
arrival at the different detectors). Indeed, from Eq. (\ref{probmes})
it results that the probability distribution of $t$ is $|g\left(M
t\right)|^2$. This immediately implies that the average time of
arrival is determined to an accuracy
\begin{eqnarray} 
\Delta t=\frac{\Delta\tau}M
\;\label{valorimedi},
\end{eqnarray}
where $\Delta\tau$ is the same of Eq. (\ref{classictime}). This result
shows a $\sqrt{M}$ improvement over the classical case
(\ref{classictime}).\par 

To emphasize the importance of
entanglement, Eq. (\ref{valorimedi}) should be compared to the result
one would obtain from an unentangled state analogous to $|\Psi\rangle_{en}$. 
To this end, consider the state defined as
\begin{eqnarray} |\Psi\rangle_{un}\equiv\bigotimes_{i=1}^M\int
d\omega_i\;\phi_{\omega_i}\;|\omega_i\rangle_i
\;\label{frequnent},
\end{eqnarray}
which describes $M$ uncorrelated single photon pulses each with
spectral function $\phi_\omega$.  By looking at the spectrum of the
state obtained by tracing away all but one of the modes in
(\ref{freqmes}), each of the photons in (\ref{frequnent}) can be shown
to have the same spectral characteristics as the photons in the
entangled state $|\Psi\rangle_{en}$. Now, using
Eq. (\ref{timeofarrivalprob}) for the uncorrelated $M$ photon pulses
$|\Psi\rangle_{un}$, it follows that
\begin{eqnarray}
p(t_1,\cdots,t_M) \propto \prod_{i=1}^M|g(t_i)|^2
\;\label{probunent},
\end{eqnarray}
which is the same result that was obtained for the classical state
(\ref{statoclassico}). Thus Eq. (\ref{classictime}) holds, with $N=1$,
also for $|\Psi\rangle_{un}$. From the comparison of
Eqs. (\ref{classictime}) and (\ref{valorimedi}), one sees that,
employing frequency-entangled pulses, an accuracy increase by a factor
$\sqrt{M}$ is obtained in the measurement of $t$ with respect to the
case of unentangled photons.\par

Since $|\Psi\rangle_{en}$ is tailored as to give the least
indetermination in the quantity $t$, it is appropriate for the
geometry of the case given in Fig. {\ref{f:localizz}}, where the sum
of the time of arrival is needed. Other entangled states can be
tailored for different geometric dispositions of the detectors, as
will be shown through some examples.\par

How is it possible to create the needed entangled states?  In the case
$M=2$, the twin beam state at the output of a cw pumped parametric
downconverter will be shown to be fit. It is a 2 photon frequency
entangled state of the form $\int
d\omega\;\phi_\omega|\omega\rangle_s|\omega_0-\omega\rangle_i$, where
$\omega_0$ is the pump frequency and $s$ and $i$ refer to the signal
and idler modes respectively.  This state is similar to
(\ref{freqmes}) and it can be employed for position measurements when
the two reference points are in opposite directions, {\it e.g.} one to
the left and one to the right of Alice. In fact, it can be seen that
$p(t_1,t_2)\propto|g(t_1-t_2)|^2$ and hence such a state is optimized
for time of arrival difference measurements, as experimentally
reported in {\cite{manou}}. In the case of $M=3$, a suitable state can
be obtained starting from a $3$-photon generation process that creates
a state of the form $\int d\omega d\omega'\;{f}(\omega,\omega')
|\omega\rangle|\omega'\rangle|\omega_0-\omega-\omega'\rangle$, and
then performing a non-demolition (or a post-selection) measurement of
the frequency difference of two of the photons. This would create a
maximally entangled $3$-photon state, tailored for the case in which
Alice has one detector on one side and two detectors on the other
side. However, for $M>2$, the creation of such frequency-entangled
states represents a continuous variable generalization of the GHZ
state, and, as such, is quite an experimental challenge.

Now turn to the use of number-squeezed states to enhance positioning.
The $N$-th excitation of a quantum system ({\it i.e.} the state
$|N\rangle$ of exactly $N$ quanta) has a de Broglie frequency $N$
times the fundamental frequency of the state. Its shorter wavelength
makes such a state appealing for positioning protocols. In this case,
the needed ``frequency state'' is given by $|N_\omega\rangle$, defined
as the state where all modes are in the vacuum except for the mode at
frequency $\omega$, which is in the Fock state $|N\rangle$. The
probability of measurement of $N$ quanta in a single pulse at times
$t_1,\cdots,t_N$ is given by Eq. (\ref{timeofarrivalprob}) with $M=1$
detectors. It is straightforward to see that, for a state of the form
$\int d\omega\;\phi_\omega\;|N_\omega\rangle$, the time of arrival
probability is given by \begin{eqnarray}
p(t_1,\cdots,t_N)\propto|g(\sum_{k=1}^Nt_k)|^2\;\label{pidit}.
\end{eqnarray}
Such a result must be compared to what one would obtain employing a
classical pulse $|\Psi\rangle_{cl}$ of $N$ mean number of photons,
{\it i.e.}  the state (\ref{statoclassico}) with $M=1$.  Its
probability (\ref{classicp}) shows that employing the $N$-photon Fock
state gives an accuracy increase of $\sqrt{N}$ {\it vs.} the coherent
state with $N$ mean number of photons. The similarity of this result
(\ref{pidit}) with the one obtained in Eq. (\ref{probmes}) stems from
the fact that the Fock state $|N_\omega\rangle$ can be interpreted as
composed by $N$ one-photon pulses of identical frequency. Hence, all
the results and considerations obtained previously apply here.  An
experiment which involves such a state for $N=2$ is reported in
{\cite{padua}}.\par

Entangled pulses of number-squeezed states combine both these
enhancements.  By replacing $|\omega\rangle$ with the number-squeezed
states $|N_\omega\rangle$ in the $M$-fold entangled state
(\ref{freqmes}), one immediately obtains
an improvement of $\sqrt{MN}$ over the accuracy obtainable by using
$M$ classical pulses of $N$ photons each.

\begin{figure}[hbt]
\begin{center}\epsfxsize=.5
\hsize\leavevmode\epsffile{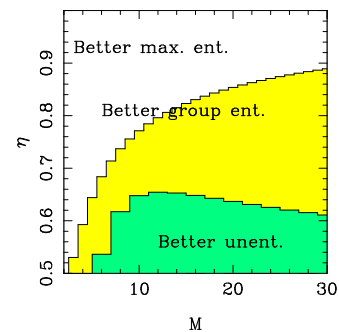}
\end{center}
\caption{Sensitivity to loss. The quantum efficiency $\eta$ needed for 
having an accuracy increase over the unentangled state
$|\Psi\rangle_{un}$ is plotted {\it vs.} the number $M$ of photons
(here $N=1$). The upper white region is where $|\Psi\rangle_{en}$ does
better than $|\Psi\rangle_{un}$. The white and light grey regions are
where a partially entangled state, which exploits a configuration
where one partially entangles subgroups of $2$ maximally entangled
photons, does better than $|\Psi\rangle_{un}$. The lower dark region
is where $|\Psi\rangle_{un}$ does better.}
\label{f:loss}\end{figure}

The enhanced accuracy achieved comes at the cost of an enhanced
sensitivity to loss.  If one or more of the photons fails to arrive,
the time of arrival of the remaining photons do not convey any timing
information.  The simplest way to solve this problem is to ignore all
trials where one or more photons is lost.  A more sophisticated method
is to use {\it partially} entangled states: these states provide a
lower level of accuracy than fully entangled states, but are more
tolerant to loss.  As shown in figure {\ref{f:loss}}, even the simple
protocol of ignoring trials with loss still surpasses the unentangled
state accuracy limit even for significant loss levels.  The use of
intrinsically loss-tolerant, partially entangled states does even
better \cite{entloss}.

Before closing, it is useful to consider the following intuitive
picture of quantum measurements of timing.  A quantum system such as a
pulse of photons or a measuring apparatus with spread in energy
$\Delta E$ can evolve from one state to an orthogonal state in time
$\Delta t$ no less than $\pi \hbar/(2 \Delta E)$ {\cite{peres}}.
Accordingly, to make more accurate timing measurements, one requires
states with sharp time dependence, and hence high spreads in energy.
Classically, combining $M$ systems each with spread in energy $\Delta
E$ results in a joint system with spread in energy $\sqrt{M}\Delta E$.
Quantum-mechanically, however, $M$ systems can be put in entangled
states in which the spread in energy is proportional to $M\Delta E$.
Similarly, $N$ photons can be joined in a squeezed state with spread
in energy $N\Delta E$.  The Margolus-Levitin theorem {\cite{margolus}}
limits the time $\Delta t$ it takes for a quantum system to evolve
from one state to an orthogonal one by $ \Delta t
\geq 2\hbar/\pi E$, where $E$ is the average energy of a system
(taking the ground state energy to be 0).  This result implies that the
$\sqrt{MN}$ enhancement presented here is the best one can do.

In conclusion, quantum entanglement and squeezing have been shown to
increase the accuracy of position measurements, and, as a consequence,
they can also be employed to improve the accuracy in distant clock
synchronization. For maximally entangled $M$-particle states we have
shown an accuracy increase $\propto\sqrt{M}$ {\it vs.} unentangled
states with identical spectral characteristics. A further increase
$\propto\sqrt{N}$ in accuracy in comparison with classical pulses was
also shown for the measurement of $N$ quanta states. At least for the
simple cases of $M=2$ or $N=2$, the described protocols are realizable
in practice.\par

This work was funded by the ARDA, NRO, and by ARO under a MURI
program.

\end{multicols}
\end{document}